\documentclass{sig-alternate-05-2015}

\begin{document}


\doi{https://hzwu.github.io/}

\isbn{:~A~2-page~summary~has~been~accepted~by~ACM~IH\&MMSec'2016}



%

\title{Prediction-error of Prediction Error (PPE)-based Reversible Data Hiding}
%
%
%
%
%

\numberofauthors{1} 
%
\author{
%
%
\alignauthor
 Han-Zhou Wu\textsuperscript{1},~Hong-Xia Wang\textsuperscript{1}~and~Yun-Qing Shi\textsuperscript{2}\\~\\
       \textsuperscript{1}\affaddr{Southwest Jiaotong University, Chengdu 611756, China}\\
	   \affaddr{h.wu.phd@ieee.org, hxwang@home.swjtu.edu.cn}\\
	   \textsuperscript{2}\affaddr{New Jersey Institute of Technology, Newark, NJ 07102, USA}\\
       \affaddr{shi@njit.edu}
}

\maketitle
\begin{abstract}
This paper presents a novel reversible data hiding (RDH) algorithm for gray-scaled images, in which the prediction-error of prediction error (PPE) of a pixel is used to carry the secret data.
In the proposed method, the pixels to be embedded are firstly predicted with their neighboring pixels to obtain the corresponding prediction errors (PEs).
Then, by exploiting the PEs of the neighboring pixels, the prediction of the PEs of the pixels can be determined. 
And, a sorting technique based on the local complexity of a pixel is used to collect the PPEs to generate an ordered PPE sequence so that, smaller PPEs will be processed first for data embedding.
By reversibly shifting the PPE histogram (PPEH) with optimized parameters, the pixels corresponding to the altered PPEH bins can be finally modified to carry the secret data.
Experimental results have implied that the proposed method can benefit from the prediction procedure of the PEs, sorting technique as well as parameters selection, and therefore 
outperform some state-of-the-art works in terms of payload-distortion performance when applied to different images.

\end{abstract}

%
%


%
%

%
%



\section{Introduction}
Reversible data hiding (RDH) \cite{ni:rev, tian:rev} aims to embed additional data such as source information into a host signal (e.g., digital image) by slightly altering the host signal, 
while ensuring that the hidden information and the host signal can be fully recovered from the marked content on the receiver side. 
Since RDH allows the original signal to be perfectly reconstructed, as a special means of information hiding, 
the RDH techniques are quite desirable and helpful in some sensitive applications such as medical image processing, 
remote sensing and military communication.

Early RDH methods \cite{celik:los, fridrich:inv} mainly use lossless compression techniques to substitute a part of the host with the compressed code and the secret data. 
These methods usually correspond to complex computation and a limited embedding capacity. 
Later on, more efficient techniques are introduced to increase the embedding capacity and/or keep the embedding distortion low, 
such as difference expansion (DE) \cite{tian:rev} and histogram shifting (HS) \cite{ni:rev}. Since the DE and HS can reduce the embedding distortion and provide a sufficient embedding payload, 
various RDH techniques have been developed along these two lines to maintain a good payload-distortion performance, e.g., 
integer wavelet transform (IWT) \cite{alattar:rev, lee:rev, wang:eff}, prediction error (PE) \cite{hong:rev, hsu:rev, li:ano, luo:rev}, prediction error expansion (PEE) \cite{chen:rev, dragoi:loc, li:eff, ou:pai, sachnev:rev, thodi:exp}, and so on. 
In addition, the RDH in terms of the payload limit subjected to a given distortion has also been studied and designed to approach the theoretical bound \cite{lin:sca, zhang:imp}.

In most cases, to achieve a better performance, the existing methods usually predict the pixels with a well-designed predictor at first. 
Then, secret bits are embedded by modifying the resultant prediction-error histogram (PEH). 
In general, a more sharply distributed PEH will provide a larger capacity or a lower distortion. 
Accordingly, the pixel prediction procedure is required to well predict the pixels. 
Due to the spatial correlations between neighboring pixels, the existing methods often exploit the PE of a pixel to carry extra data. 
Actually, there should also exist strong correlations between neighboring PEs. 
One evidence can be found in the prediction mechanism of modern video lossy compression. 
For example, in intra prediction, the prediction block for an intra $4\times 4$ luma macroblock (consisting of 16 pixels) can be generated with nine 
possible prediction modes due to the spatial correlations between neighboring pixels. Then, in order to improve the coding efficiency, 
the prediction mode of a luma macroblock should be further predicted from the prediction modes of neighboring luma macroblocks since there also exists 
strong correlations between neighboring prediction modes. The success of steganalysis by modeling the differences between 
neighboring pixels with first-order and second-order Markov chains \cite{pev:analysis} also reveals that there exists correlations between neighboring PEs if we consider 
the differences as a kind of prediction errors.

Therefore, instead of directly exploiting the PE of a pixel, 
we can actually adopt the prediction-error of prediction error (PPE) of a pixel to embed secret data.
Based on this perspective, we are to propose a reliable PPE-based RDH method in this paper, 
in which the PPEs of the pixels to be embedded are firstly determined according to their neighboring pixels. 
To reduce the embedding distortion for a required payload, the pixels are then sorted by their local complexities. 
A pixel with a lower complexity will be preferentially embedded with a secret bit. By shifting the resultant PPE histogram (PPEH), 
the corresponding pixels can be finally modified to accommodate the secret data. Experimental results have implied that, 
the proposed method can benefit from the obtained PPEH and sorting procedure, and therefore provide a good payload-distortion behavior.

The rest of this paper are organized as follows. 
Section 2 introduces the detailed RDH algorithm. 
In Section 3, we show our experiments and analysis to evaluate the proposed method. 
Finally, we conclude this paper in Section 4.

\begin{figure}
\centering
\includegraphics[height=1.6in, width=3.4in]{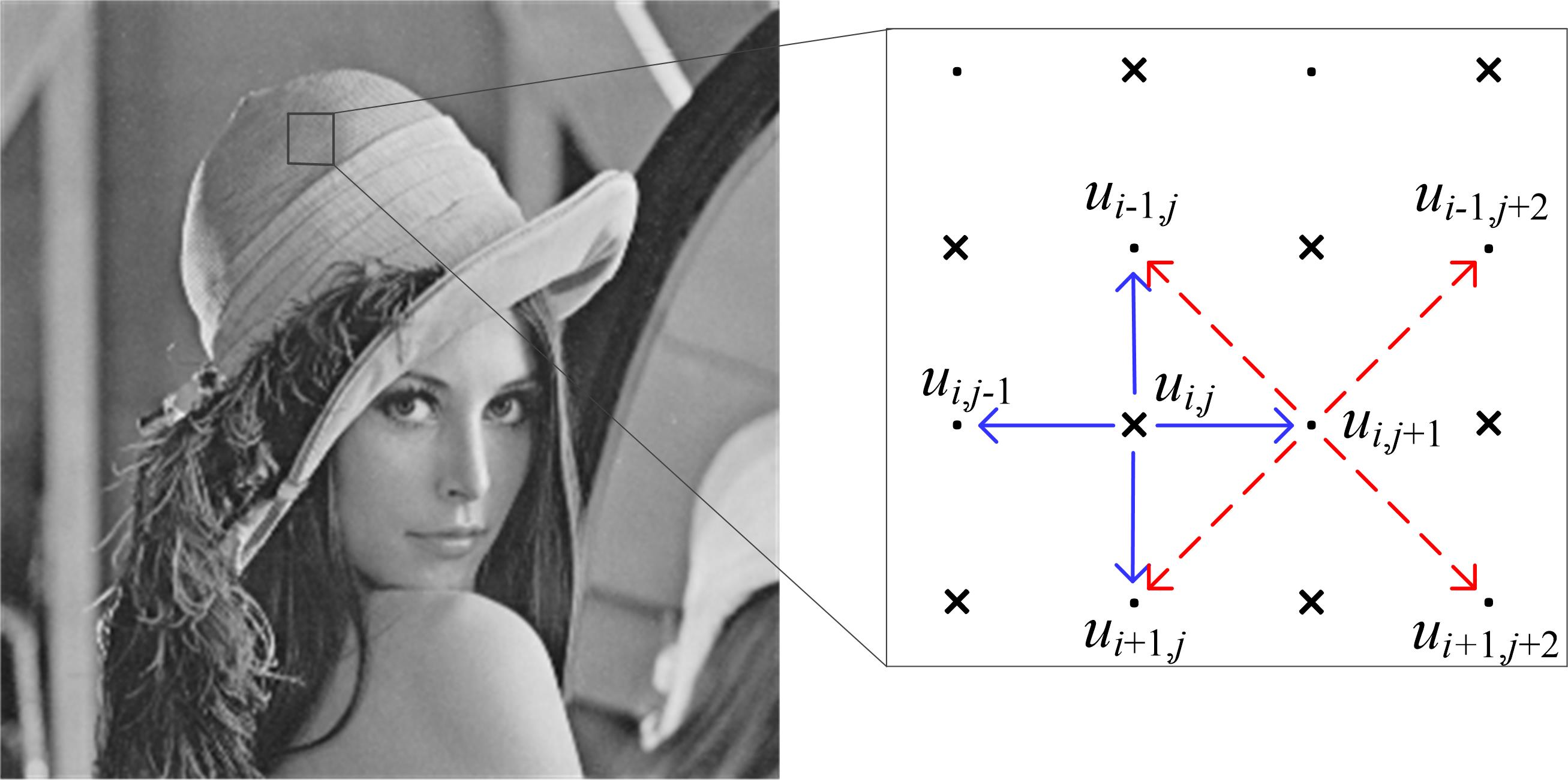}
\caption{The pixel prediction pattern. The pixel $u_{i,j}$ will be predicted from its neighboring pixels in the dot set, 
and the pixel $u_{i,j\text{+1}}$ will be predicted from its neighboring pixels still in the dot set.}
\end{figure}

\section{Proposed Method}
In this section, we give a detailed introduction about the proposed algorithm with the following two aspects: 1) data embedding; 
and 2) data extraction and image recovery.

\subsection{Data Embedding}
The proposed method uses gray-scaled images. 
The data embedding procedure mainly consists of four parts: the to-be-embedded pixel prediction, 
the prediction of prediction error, the use of pixel sorting, and the data hiding in the image. 
We introduce each part in detail in the following.

\subsubsection{Pixel Prediction}
For data embedding, the pixels to be embedded are predicted from their neighboring pixels, by which the corresponding PEs can be collected. 
Specifically, all image pixels are divided into two sets: the cross set and dot set, as shown in Fig. 1. 
The cross set is used for data embedding and dot set for pixel prediction. 
We use this rhombus pattern as it can maintain a good prediction performance \cite{ou:pai, sachnev:rev}. 
It is noted that, one may use other efficient prediction patterns. 
In the following, we consider the pixel $u_{i,j}$ in Fig. 1, to introduce the to-be-embedded pixel prediction procedure.

At first, $u_{i,j}$ will be predicted from its four neighboring pixels in the dot set by applying interpolation operation \cite{luo:rev}. 
we predict $u_{i,j}$ along the horizontal and vertical directions, respectively. The two directional predictors are defined as:
\begin{equation}
{u_{i,j}}^{'}=\frac{u_{i,j-1}+u_{i,j+1}}{2},~{u_{i,j}}^{''}=\frac{u_{i-1,j}+u_{i+1,j}}{2}.
\end{equation}

We adopt the weighted average between ${u_{i,j}}^{'}$ and ${u_{i,j}}^{''}$ to obtain the prediction value of $u_{i,j}$, which is computed as:
\begin{equation}
{u_{i,j}}^{+}=\text{Round}(w_{i,j}\cdot {u_{i,j}}^{'} + (1-w_{i,j})\cdot {u_{i,j}}^{''}).
\end{equation}
Here, Round($x$) denotes a function that returns the nearest integer to $x$, and; $w_{i,j}$ is defined as:
\begin{equation}
w_{i,j} = \frac{{\sigma_{i,j}}^{''}}{{\sigma_{i,j}}^{'}+{\sigma_{i,j}}^{''}}.
\end{equation}
where
\begin{equation}
{\sigma_{i,j}}^{'} = \frac{1}{3}\cdot \sum_{v\in\{u_{i,j-1}, {u_{i,j}}^{'}, u_{i,j+1}\}} (v-\frac{{u_{i,j}}^{'} + {u_{i,j}}^{''}}{2})^2.
\end{equation}
\begin{equation}
{\sigma_{i,j}}^{''} = \frac{1}{3}\cdot \sum_{v\in\{u_{i-1,j}, {u_{i,j}}^{''}, u_{i+1,j}\}} (v-\frac{{u_{i,j}}^{'} + {u_{i,j}}^{''}}{2})^2.
\end{equation}

In this way, we can determine out the prediction value of the pixel $u_{i,j}$. It is straightforward to process other pixels in the cross set with the similar procedure.

\subsubsection{Prediction of Prediction Error}
After predicting $u_{i,j}$, the resultant prediction-error can be obtained as $e_{i,j}=u_{i,j}-{u_{i,j}}^{+}$. 
The prediction of $e_{i,j}$, denoted by ${e_{i,j}}^{'}$, will be computed from the PEs of neighboring pixels. 
Let $e_{i-1,j}$, $e_{i,j+1}$, $e_{i+1,j}$ and $e_{i,j-1}$ denote the PEs of $u_{i-1,j}$, $u_{i,j+1}$, $u_{i+1,j}$ and $u_{i,j-1}$, respectively. 
${e_{i,j}}^{'}$ will be defined as:
\begin{equation}
{e_{i,j}}^{'}=\text{Round}(\frac{e_{i-1,j}+e_{i,j+1}+e_{i+1,j}+e_{i,j-1}}{4}).
\end{equation}

Unlike $u_{i,j}$, the neighboring pixels $u_{i-1,j}$, $u_{i,j+1}$, $u_{i+1,j}$ and $u_{i,j-1}$ will be all predicted along two diagonal directions (see Fig. 1), respectively. 
It indicates that, these pixels will be predicted from their neighboring pixels still in the dot set, which can ensure reversibility. 
In the following, we present the method to predict the pixel $u_{i,j+1}$ in Fig. 1 with its neighboring pixels $u_{i-1,j}$, $u_{i-1,j+2}$, $u_{i+1,j+2}$, and $u_{i+1,j}$. 

At first, we predict $u_{i,j+1}$ along the two diagonal directions, respectively, denoted by:
\begin{equation}
{u_{i,j+1}}^{'}=\frac{u_{i-1,j}+u_{i+1,j+2}}{2},~{u_{i,j+1}}^{''}=\frac{u_{i-1,j+2}+u_{i+1,j}}{2}.
\end{equation}

Similarly, we use the weighted average between ${u_{i,j+1}}^{'}$ and ${u_{i,j+1}}^{''}$ to 
obtain the final prediction value of $u_{i,j+1}$, which can be computed as:
\begin{equation}
{u_{i,j+1}}^{+}=\text{Round}(w_{i,j+1}\cdot {u_{i,j+1}}^{'} + (1-w_{i,j+1})\cdot {u_{i,j+1}}^{''}).
\end{equation}
Here, $w_{i,j+1}$ is determined by:
\begin{equation}
w_{i,j+1} = \frac{{\sigma_{i,j+1}}^{''}}{{\sigma_{i,j+1}}^{'}+{\sigma_{i,j+1}}^{''}}.
\end{equation}
where
\begin{equation}
{\sigma_{i,j+1}}^{'} = \frac{1}{3}\cdot \sum_{v\in\{u_{i-1,j}, {u_{i,j+1}}^{'}, u_{i+1,j+2}\}} (v-u_{i,j+1})^2.
\end{equation}
\begin{equation}
{\sigma_{i,j+1}}^{''} = \frac{1}{3}\cdot \sum_{v\in\{u_{i-1,j+2}, {u_{i,j+1}}^{''}, u_{i+1,j}\}} (v-u_{i,j+1})^2.
\end{equation}

Therefore, we can obtain the PE of the pixel $u_{i,j+1}$, i.e., $e_{i,j+1}=u_{i,j+1}-{u_{i,j+1}}^{+}$. 
It is noted that, in Eq. (10, 11), the original value of $u_{i,j+1}$ is used since the pixels in the dot set are kept unchanged during data embedding. 
Obviously, the prediction procedure can be easily applied to the other pixels $u_{i-1,j}$, $u_{i+1,j}$ and $u_{i,j-1}$. 
Thus, we can finally compute the prediction of $e_{i,j}$ as Eq. (6), and, 
the prediction-error of $e_{i,j}$ can be determined as ${e_{i,j}}^{+}=e_{i,j}-{e_{i,j}}^{'}$.

\subsubsection{Pixel Sorting}
In addition to the prediction of PEs, a pixel sorting (also called pixel selection) technique motivated by previous works \cite{ou:pai, sachnev:rev} 
is used in the proposed method. The use of sorting is to put the PPEs in a decreasing order of prediction accuracy so that smaller PPEs can be processed first. 
The pixels to be embedded are sorted according to their local complexities. The local complexity for $u_{i,j}$ is defined as:
\begin{equation}
\epsilon_{i,j}=\left [ \frac{1}{6}\cdot \sum_{s=1}^{6}({\varrho_{i,j}}^{(s)}-\sum_{t=1}^{6}{\varrho_{i,j}}^{(t)}/6)^2 \right ]^{1/2}.
\end{equation}
where ${\varrho_{i,j}}^{(1)}=|u_{i-1,j}-u_{i,j+1}|$, 
${\varrho_{i,j}}^{(2)}=|u_{i-1,j}-u_{i+1,j}|$, ${\varrho_{i,j}}^{(3)}=|u_{i-1,j}-u_{i,j-1}|$, 
${\varrho_{i,j}}^{(4)}=|u_{i,j+1}-u_{i+1,j}|$, ${\varrho_{i,j}}^{(5)}=|u_{i,j+1}-u_{i,j-1}|$, and 
${\varrho_{i,j}}^{(6)}=|u_{i+1,j}-u_{i,j-1}|$.

In general, a lower local complexity indicates that, the pixel can be predicted with a relatively lower PPE. 
It should be noted that, one may use other efficient methods to evaluate the local complexity of a pixel. 
Therefore, according to the local complexities, the corresponding pixels can be sorted in a decreasing order of the prediction accuracy.

\subsubsection{Data Hiding with PPEs}
After sorting, we can obtain an ordered pixel-sequence as well as the corresponding PPE sequence. 
It implies that, we could collect a part of the PPEs to generate a PPEH, which will follow a Laplacian-like distribution centered at the zero-bin. 
The secret data will be embedded into the given image by modifying the pixels corresponding to the shifted PPEH bins.
Clearly, for a fixed PPEH, we choose two peak-zero bin-pairs, denoted by ($l_p$, $l_z$) and ($r_p$, $r_z$), for data embedding. 
Without the loss of generality, we assume that, $l_z<l_p<r_p<r_z$, $h(l_p)>0$, $h(r_p)>0$, and $h(l_z)=h(r_z)=0$. Here, $h(x)$ denotes the amount of occurrences of the bin with a value of $x$.
During the data embedding, the PPEs located in range $(-\infty,l_z)\cup(l_p,r_p)\cup(r_z,+\infty)$ will be kept unchanged. 
The PPEs located in range $[l_z,l_p)\cup(r_p,r_z]$ will be shifted to avoid ambiguous. 
And, for a PPE with a value of $l_p$ or $r_p$, if the secret bit equals ``0'', the PPE will be kept unchanged; 
otherwise, it will be shifted along the corresponding direction to carry ``1''. 
Thus, an altered PPEH can be obtained, and the used pixels can be finally modified with \{+1, 0, -1\} operation to match the PPEH. 
We take the pixel $u_{i,j}$ in Fig. 1 for example. At first, the marked PPE of $u_{i,j}$ during data embedding is computed as:
\begin{equation}
{e_{i,j}}^{*}=\left\{\begin{matrix}
{e_{i,j}}^{+}+\text{sgn}({e_{i,j}}^{+}-\frac{l_p+r_p}{2})\cdot b,\text{if}~{e_{i,j}}^{+}\in\{l_p, r_p\};~~~~~~~\\
{e_{i,j}}^{+}+\text{sgn}({e_{i,j}}^{+}-\frac{l_p+r_p}{2}),~\text{if}~{e_{i,j}}^{+}\in[l_z,l_p)\cup(r_p,r_z];\\
{e_{i,j}}^{+},~\text{otherwise}.~~~~~~~~~~~~~~~~~~~~~~~~~~~~~~~~~~~~~~~~~~~~~~~~
\end{matrix}\right.
\end{equation}
where $b\in\{0, 1\}$ is the secret bit to be embedded, and $\text{sgn}(t)$ denotes a function that returns $t/|t|$.
 
Then, the marked version of $u_{i,j}$ will be determined as:
\begin{equation}
{u_{i,j}}^{*}={u_{i,j}}^{+}+{e_{i,j}}^{'}+{e_{i,j}}^{*}.
\end{equation}

In this way, the secret data can be successfully embedded. 
It is noted that, there is no need that $h(l_p)+h(r_p)$ is maximal as long as $h(l_p)+h(r_p)$ is larger than the size of the required payload 
to be embedded. And, it would be desirable that, the expected number of modified pixels is as small as possible 
so as to keep the introduced distortion low since a larger number of modified pixels usually corresponds to a higher distortion. 
Since altering a pixel during data embedding may result in overflow/underflow problem, to avoid this, 
the boundary pixels in the cross set should be shifted in advance and recorded to produce a location map, 
which will be embedded together with the secret data. Since the boundary pixels in nature images are relatively rare, 
the effect on the pure embedding payload could be ignored. 
Based on the above analysis, we are now ready to describe the proposed PPE-based RDH procedure as follows.

\textbf{Step 1)} Empty the LSBs of some specified pixels in the dot set to store the data embedding parameters such as the data hiding key, 
$l_p$, $l_z$, $r_p$, and $r_z$; and, then take the replaced LSBs as a part of the to-be-embedded data. 

\textbf{Step 2)} Adjust the boundary-pixels in the cross set into the reliable range; record their values and positions, 
and thus construct a location map, which will be losslessly compressed and also constitute a part of the to-be-embedded data.

\textbf{Step 3)} Utilize the pixels in the cross set to finally construct an ordered pixel-sequence as well as the corresponding PPE sequence, 
according to above-mentioned pixel prediction, prediction of prediction error, and sorting techniques.

\textbf{Step 4)} Orderly select an unprocessed PPE from the PPE sequence, and modify the corresponding pixel according to Eq. (13, 14). 
Repeat the current step until the entire to-be-embedded data (consists of the replaced LSBs, the compressed location map, 
and the secret data) are embedded.

\textbf{Step 5)} According to an inverse operation, use the modified pixels to finally form the marked image.

\subsection{Data Extraction and Image Recovery}
After the data receiver acquires the marked image, he can completely extract the hidden information as well as recover the original image without loss. 
It can be performed with an inverse operation on the data hider side. We consider the pixel $u_{i,j}$ in Fig. 1 as an example. 
Obviously, at first, the data receiver can correctly determine ${u_{i,j}}^{+}$ and ${e_{i,j}}^{'}$ since the pixels in the dot set were kept unchanged. 
Then, the marked PPE can be easily reconstructed as:
\begin{equation}
{e_{i,j}}^{*}={u_{i,j}}^{*}-{u_{i,j}}^{+}-{e_{i,j}}^{'}.
\end{equation}

Accordingly, a secret bit ``0'' can be extracted if the marked PPE equals $l_p$ or $r_p$; 
otherwise, a secret bit ``1'' can be obtained if it equals $l_p-1$ or $r_p+1$. And, the original PPE ${e_{i,j}}^{+}$ can be recovered as:
\begin{equation}
{e_{i,j}}^{+}=\left\{\begin{matrix}
{e_{i,j}}^{*}-\text{sgn}({e_{i,j}}^{*}-\frac{l_p+r_p}{2}),~\text{if}~{e_{i,j}}^{*}\in[l_z,l_p)\cup(r_p,r_z];\\ 
{e_{i,j}}^{*},~\text{otherwise}.~~~~~~~~~~~~~~~~~~~~~~~~~~~~~~~~~~~~~~~~~~~~~~
\end{matrix}\right.
\end{equation}

Therefore, the original pixel $u_{i,j}$ can be perfectly reconstructed as:
\begin{equation}
u_{i,j}={u_{i,j}}^{+}+{e_{i,j}}^{'}+{e_{i,j}}^{+}.
\end{equation}

In this way, the hidden information as well as the original image can be fully reconstructed. 
Based on the above analysis, the procedure of data extraction and image recovery can be described as follows.

\textbf{Step 1)} Extract the LSBs of the specified pixels to determine the data embedding parameters.

\textbf{Step 2)} Utilize the pixels in the cross set to finally construct an ordered pixel-sequence as well as the corresponding PPE sequence 
with the same procedure on the sender side.

\textbf{Step 3)} Orderly select an unprocessed PPE from the PPE sequence, exact the hidden information and recover the corresponding pixel according to Eq. (15, 16, 17). 
Repeat the current step until the entire hidden data are fully extracted.

\textbf{Step 4)} Retrieve the original secret message from the extracted data; and, further recover the original image without error 
according to the extracted LSBs and location map.

Noting that, changes in the cross set will not affect the dot set. 
It indicates that, instead of only the cross set, the dot set can be applied for data embedding after data embedding with the cross set. 
The advantage is that, when using only the cross set, pixels with larger PPEs have to be modified so as to 
carry the required payload; while, for the consecutive usage of the cross set and dot set, two sets of sorted PPEs with smaller values can be used first, 
which implies that, the required payload of each set is approximately half of that for data embedding only with the cross set, and thus could maintain a lower distortion.

\section{Experiments and analysis}
\subsection{Parameters Selection}
For data embedding, we need to select suitable values for $(l_p, l_z)$ and $(r_p, r_z)$.
It is required that, $(l_p, l_z)$ and $(r_p, r_z)$ should be able to carry the required payload, and it is 
desirable to keep the introduced distortion as low as possible. We here present an approximation algorithm to 
select suitable $(l_p, l_z)$ and $(r_p, r_z)$, by minimizing the amount of possibly-shifted pixels. The reason is that,
the introduced distortion relies on the shifting operation on the PPEH, and a larger number of shifted pixels usually 
corresponds to a higher distortion. We often expect to reduce the number of shifted pixels so as to keep the distortion low.
In our parameters selection algorithm, though the selected $(l_p, l_z)$ and $(r_p, r_z)$ may be not optimal, our experiments have 
shown that, the selected $(l_p, l_z)$ and $(r_p, r_z)$ can be determined with a low computational complexity, and provide a good 
payload-distortion performance. In the following, we introduce the procedure to select $(l_p, l_z)$ and $(r_p, r_z)$ in detail.

At first, for a usable PPEH, there should exist such two PPEH bins $(x,y)$ that $h(x)+h(y)\geq\rho$, where $\rho$ is 
the size of the entire payload to be embedded. In applications, $\rho$ can be determined or roughly estimated. 
Under this condition, we compute $l_z$ and $r_z$ as:
\begin{equation}
l_z=\underset{x<0,~h(x)=0}{\text{arg min}}~|x|,~r_z=\underset{x>1,~h(x)=0}{\text{arg min}}~|x|.
\end{equation}

We compute $l_p$ and $r_p$ by minimizing an approximation of the amount of possibly-shifted pixels, which is defined as:
\begin{equation}
(l_p, r_p)=\underset{l_z < x_l < x_r < r_z,~h(x_l)+h(x_r)\geq \rho}{\text{arg min}}~\frac{\rho}{2}+\sum_{k\in[l_z, x_l)\cup(x_r, r_z]}h(k).
\end{equation}

Here, the first item is the expected number of possibly-shifted pixels for data hiding, and the second item 
is the approximation value of possibly-shifted pixels for reversibility. Therefore, we can find $(l_p, l_z)$ and $(r_p, r_z)$ 
for a usable PPEH. It can be observed that, the complexity to compute $(l_p, r_p)$ is $O(|l_z-r_z|^2)$. 
In applications, $|l_z|$ and $|r_z|$ are both small, indicating that, the complexity is very low.

For a required payload, since we may only use a part of the PPEs for data hiding, there may exist different usable PPEHs.
It means that, we need to further choose a suitable PPEH from an ordered PPE sequence. As shown in \textbf{Algorithm 1}, we propose to 
find the suitable PPEH by orderly collecting the PPEs with a step size $L$. Here, it is noted that, the PPE sequence shown in Algorithm 1 is obtained by without embedding the data-embedding parameters, which indicates that, 
it may be not exactly the same as the actual PPE sequence to be embedded.
However, it will be rather close to the actual one since the bit-size of the data embedding parameters is rather small and thus its impact on the generation of the PPE sequence can be ignored.
It can be seen that, we need to process at most $\left \lceil T/L  \right \rceil$ PPEH(s). In applications, the computational complexity would be low 
by setting a suitable $L$, e.g., $L = \rho/2,~\rho$, etc. Therefore, by applying the proposed approximation algorithm, we can find two suitable
peak-zero bin-pairs for data embedding. \\
-----------------------------------------------------------------------------\\
\textbf{Algorithm 1:} Proposed Parameters Selection Alogrithhm\\
\textbf{Input:} $\rho$: the size of entire payload to be embedded;\\
$~~~~~~~~~~~~$$[{e_1}^{+}, {e_2}^{+}, ..., {e_T}^{+}]$: an ordered PPE sequence;\\
$~~~~~~~~~~~~$$T$: the total number of PPEs; $L$: the step size.\\ 
\textbf{Output:} $(l_p, l_z)$, $(r_p, r_z)$: two peak-zero bin-pairs.\\
1: \textbf{Procedure} $ParametersSelection$\\
2:~~~~~\textbf{Set} $h(x)\leftarrow 0$ for all possible $x$\\
3:~~~~~\textbf{For} $k=1,2,...,T$ \textbf{do}\\
4:~~~~~~~~~~\textbf{Set} $h({e_k}^{+})\leftarrow h({e_k}^{+})+1$\\
5:~~~~~~~~~~\textbf{If} $L~|~k$ or $k~=~T$ \textbf{do}\\
6:~~~~~~~~~~~~~~~\textbf{If} exists such $(x,y)$ that $h(x)+h(y)\geq \rho$ \textbf{do}\\
7:~~~~~~~~~~~~~~~~~~~~\textbf{Find} $(l_p, l_z)$, $(r_p, r_z)$ with Eq. (18, 19)\\
8:~~~~~~~~~~~~~~~~~~~~\textbf{Return} $(l_p, l_z)$, $(r_p, r_z)$\\
9:~~~~~\textbf{Return} Failure\\
-----------------------------------------------------------------------------

\subsection{Payload-Distortion Performance}
We have implemented the proposed PPE-based RDH algorithm using Microsoft Visual C++ and CxImage Library\footnote{http://www.xdp.it/cximage.htm}, 
and applied it to the standard testing images (sized 512$\times$512, and 8-bit gray-scaled): Airplane, Lena, Baboon, and (Fishing) Boat, in which 
the equivalence between the reconstructed image and original one has proved that our RDH algorithm maintains reversibility.

\begin{figure*}[t!]
\centering
\includegraphics[height=6.4in, width=6.4in]{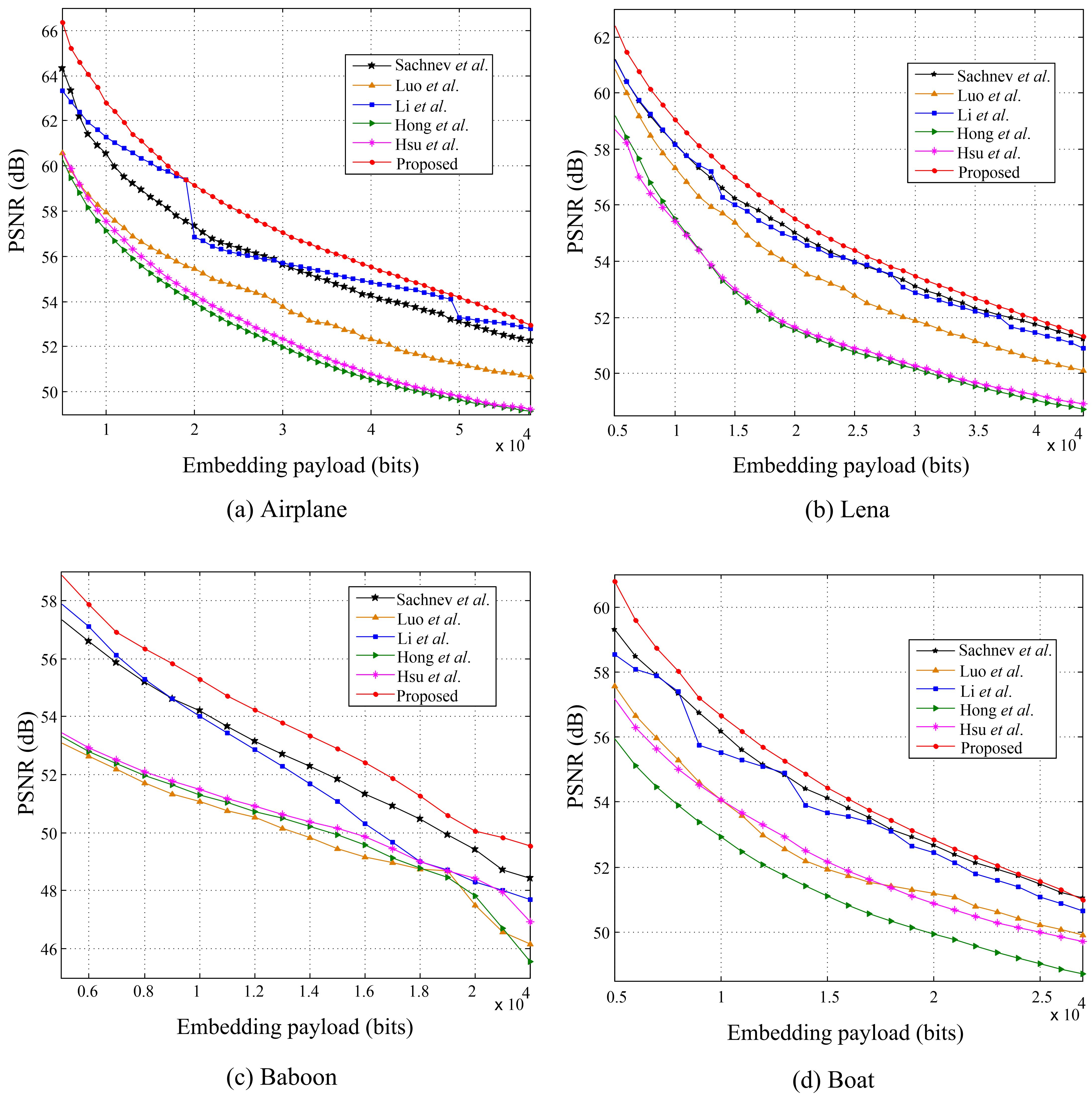}
\caption{The payload-distortion performance comparison between the state-of-the-art methods of 
Sachnev \emph{et al}. \cite{sachnev:rev}, Hong \emph{et al}. \cite{hong:rev}, Luo \emph{et al}. \cite{luo:rev}, 
Li \emph{et al}. \cite{li:eff}, Hsu \emph{et al}. \cite{hsu:rev} and the proposed method.}
\end{figure*}

In the experiments, we employ a pseudo-random bit-string as the secret message, 
and adopt the peak signal-to-noise ratio (PSNR) value as the distortion measurement.
And, we use the payload-distortion behavior to demonstrate the performance of the proposed method.
The payload-distortion performance of the proposed method is evaluated by comparing it with the state-of-the-art 
methods of Sachnev \emph{et al}. \cite{sachnev:rev}, Hong \emph{et al}. \cite{hong:rev}, Luo \emph{et al}. \cite{luo:rev}, 
Li \emph{et al}. \cite{li:eff}, and Hsu \emph{et al}. \cite{hsu:rev}. It can be observed from Fig. 2 that, 
the proposed method outperforms these state-of-the-art works in terms of the payload-distortion performance. 
Table 1 and Table 2 show the PSNRs due to different RDH methods for the required payloads 10,000 bits and 20,000 bits, respectively.
It can be seen that, the average gains in terms of the PSNR of the proposed method are at least 1.16 dB and 0.78 dB, respectively, 
which can demonstrate the superiority of the proposed method.

\section{Conclusion and Discussion}
In this paper, we present an RDH method for gray-scaled images, 
which aims to use the prediction-error of prediction error (PPE) of a pixel to carry the message bit. 
During the data embedding, the interpolation operation is used to well predict the pixels and their PEs, and;
the pixel sorting is used for better exploiting the PPEs with small values. An approximation algorithm is proposed to 
select suitable PPEH shifting parameters so as to keep the number of to-be-shifted pixels low, which can help reduce the distortion.
Our experimental results show that the presented method works well in terms of payload-distortion performance when compared to some state-of-the-art methods. 
In the future, there is still room for investment or improvement such as by designing a better predictor (e.g., least square predictor \cite{dragoi:loc}), better evaluating local complexities, and 
applying better data hiding operation (e.g., PEE \cite{sachnev:rev, thodi:exp}, pairwise PEE \cite{ou:pai}). 
Moreover, it is possible to extend the PPE-based technique to different cover media, and/or 
employ high-order prediction errors for data embedding.

\begin{table*}
\centering
\caption{Comparisons in terms of PSNR (dB) for a required payload of 10,000 bits.}
\begin{tabular}{c|c|c|c|c|c|c} \hline
Images&Sachnev \emph{et al}.&Hong \emph{et al}.&Luo \emph{et al}.&Li \emph{et al}.&Hsu \emph{et al}.&Proposed\\ \hline
Airplane & 60.56 & 57.14 & 57.97 & 61.27 & 57.56 & \textbf{62.81}\\ \hline
Lena & 58.22 & 55.52 & 57.35 & 58.17 & 55.41 & \textbf{59.05}\\ \hline
Baboon & 54.19 & 51.29 & 51.08 & 54.02 & 51.48 & \textbf{55.27}\\ \hline
Boat & 56.17 & 52.93 & 54.07 & 55.51 & 54.08 & \textbf{56.66}\\ \hline
Average & 57.29 & 54.22 & 55.12 & 57.24 & 54.63 & \textbf{58.45}\\
\hline\end{tabular}
\end{table*}

\begin{table*}
\centering
\caption{Comparisons in terms of PSNR (dB) for a required payload of 20,000 bits.}
\begin{tabular}{c|c|c|c|c|c|c} \hline
Images&Sachnev \emph{et al}.&Hong \emph{et al}.&Luo \emph{et al}.&Li \emph{et al}.&Hsu \emph{et al}.&Proposed\\ \hline
Airplane & 57.33 & 53.94 & 55.44 & 56.84 & 54.31 & \textbf{59.16}\\ \hline
Lena & 55.06 & 51.54 & 53.85 & 54.83 & 51.65 & \textbf{55.53}\\ \hline
Baboon & 49.42 & 47.83 & 47.50 & 48.29 & 48.43 & \textbf{50.06}\\ \hline
Boat & 52.68 & 49.95 & 51.19 & 52.44 & 50.88 & \textbf{52.83}\\ \hline
Average & 53.62 & 50.82 & 52.00 & 53.10 & 51.32 & \textbf{54.40}\\
\hline\end{tabular}
\end{table*}

\section{Acknowledgment}
This work was supported by the Chinese Scholarship Council (CSC) under the grant No. 201407000030, and partially supported by the National Natural Science Foundation of China (NSFC) under the grant No. U1536110.
\bibliographystyle{abbrv}
\bibliography{sigproc}  
%
%
\end{document}